\documentclass[conference]{IEEEtran}
\usepackage{cite}
\usepackage{url}
\usepackage{amsmath,amssymb,amsfonts}
\usepackage{algorithmic}
\usepackage{graphicx}
\usepackage{textcomp}
\usepackage{xcolor}
\usepackage[inkscapelatex=false]{svg}
\usepackage{listings}
\usepackage{xcolor}
\usepackage{orcidlink}
\usepackage{tikz}
\usepackage{fancyhdr}
\usepackage{hyperref}
\hypersetup{pdfborder={0 0 0}}
\def\BibTeX{{\rm B\kern-.05em{\sc i\kern-.025em b}\kern-.08em
    T\kern-.1667em\lower.7ex\hbox{E}\kern-.125emX}}

\lstdefinelanguage{oml}{
  morekeywords={instance, concept, relation, entity, aspect, scalar, property, from, to, forward, reverse, domain, range, functional},
  sensitive=true,
  alsoletter={:,_,-},
  morecomment=[l]{//},
  morestring=[b]",
  morekeywords=[2]{DTDFVocab:Service, DTDFVocab:Enabler, DTDFVocab:Model, DTDFVocab:Data, DTDFVocab:Input, DTDFVocab:Standardization, Service, ProvidedThing,  Enabler, Model, Data, Input, Standardization, DataTransmitted, DTComponent, TimeScaleThing, base:DescribedThing},
  morekeywords=[3]{DescribedThing, IsCommEnabler, base:desc, xsd:boolean, desc, enables, enabledBy, inputTo, hasInput, DTDFVocab:enables, DTDFVocab:enabledBy, DTDFVocab:inputTo, DTDFVocab:hasInput, provides, providedBy, DTDFVocab:provides, DTDFVocab:atStage, asData, fromData},
  morekeywords=[4]{<, [, ], :},
  literate={,}{{,}}{1} 
}

\definecolor{lightblue}{RGB}{66,133,244}

\newcommand{\preprintbanner}{%
    \begin{tikzpicture}[remember picture, overlay]
        \node[rectangle, text=red!80!white, inner sep=8pt,
              anchor=north, text width=\paperwidth, align=center] 
              at ([yshift=-0.5cm]current page.north) 
              {Author pre-print. Publication accepted for \href{https://conf.researchr.org/home/edtconf-2025}{\textcolor{lightblue}{EDTconf 2025}}.};
    \end{tikzpicture}%
}

\newcommand{\preprintbannerbottom}{%
    \begin{tikzpicture}[remember picture, overlay]
        \node[rectangle, text=red!80!white, inner sep=8pt,
              anchor=south, text width=\textwidth, align=center] 
              at ([yshift=0.5cm]current page.south) 
              {\scriptsize
              © 2025 IEEE. Personal use of this material is permitted. Permission from IEEE must be obtained for all other uses, in any current or future media, including reprinting/republishing this material for advertising or promotional purposes, creating new collective works, for resale or redistribution to servers or lists, or reuse of any copyrighted component of this work in other works.
              
              \vspace{0.3em}};
    \end{tikzpicture}%
}

\begin{document}

\title{DTInsight: A Tool for Explicit, Interactive, and Continuous Digital Twin Reporting}

\author{
    \IEEEauthorblockN{K\'{e}rian Fiter\orcidlink{0009-0001-7731-0299}}
    \IEEEauthorblockA{\textit{Dept. of Computer and Software Eng.} \\
    \textit{Polytechnique Montr\'{e}al}\\
    Montr\'{e}al, Canada \\
    \url{kerian.fiter@polymtl.ca}}
\and
    \IEEEauthorblockN{Louis Malassign\'{e}-Onfroy\orcidlink{0009-0000-8634-3909}}
    \IEEEauthorblockA{
    %\textit{\'{E}cole d'Ing\'{e}nieurs du Conservatoire National\newline des Arts et M\'{e}tiers (EICNAM)}\\
    \textit{\'{E}cole d'Ing\'{e}nieurs du Conservatoire}\\\textit{National des Arts et M\'{e}tiers (EICNAM)}\\
    Paris, France \\
    \url{louis.malassigne@gmail.com}}
\and
    \IEEEauthorblockN{Bentley Oakes\orcidlink{0000-0001-7558-1434}}
    \IEEEauthorblockA{\textit{Dept. of Computer and Software Eng.} \\
    \textit{Polytechnique Montr\'{e}al}\\
    Montr\'{e}al, Canada \\
    \url{bentley.oakes@polymtl.ca}}
}

\maketitle

\preprintbanner
\preprintbannerbottom

\begin{abstract}

With Digital Twin (DT) construction and evolution occurring over time, stakeholders require tools to understand the current characteristics and conceptual architecture of the system at any time. We introduce DTInsight, a systematic and automated tool and methodology for producing continuous reporting for DTs. DTInsight offers three key features: (a) an interactive conceptual architecture visualization of DTs; (b) generation of summaries of DT characteristics based on ontological data; and (c) integration of these outputs into a reporting page within a continuous integration and continuous deployment (CI/CD) pipeline. Given a modeled description of the DT aligning to our DT Description Framework (DTDF), DTInsight enables up-to-date and detailed reports for enhanced stakeholder understanding.
\end{abstract}


\section{Introduction}

Digital Twins (DTs) are digital representations of physical systems, enabling monitoring, insight generation, and control of their physical counterparts~\cite{madni2019leveraging}. 
DT engineering is becoming increasingly systematized through model-based approaches~\cite{michaelModelDrivenEngineeringDigital2025}. However, previous work has shown that the \textit{reporting} of DT capabilities often leaves out important details~\cite{Oakes2021improving,gunasekaran2025behavioralanalysis}. We argue this lack of systematic reporting hinders research into DT capabilities and engineering process.

In particular, the motivation behind DTInsight is to apply TwinOps \cite{huguesTwinOpsDevOpsMeets2020} to DT engineering. That is, the key problem to address is: \textit{as the DT changes, how can reporting be kept up-to-date such that stakeholders (including non-technical personnel such as management) are always aware of the DT's current capabilities, architecture, and state}. We aim to improve users' understanding of their DT and enhance their decision-making capabilities regarding the DT's evolution.

Another aspect of this tool is to \textit{assist practitioners in reporting their DT}~\cite{Oakes2021improving}. This approach works towards the vision of DTs having \textit{characteristic cards} unifying their digital and physical element reporting, similar to machine learning model cards \cite{ozoani2022modelcard,Toma2025ModelCards} or a \textit{Digital Twin Bill of Materials} (DT BOM).

Our research treats DTs as a complex software system, therefore we bring in concepts from fields related to software engineering and adopt principles of \textit{observability}~\cite{majors2022observability}, to have a deeper understanding of the DT.

%Our research brings in concepts from the broader field of \textit{observability engineering}, highlighting an often-overlooked dimension in the development of DTs. While metrics and reporting tools are valuable, they typically support \textit{reactive practices}. In contrast, we advocate for a cultural shift in DT engineering toward \textit{proactive exploration}. 
This means working towards giving the user the ability to investigate and understand the DT's current structural and behavioral state. For example, our architectural view promotes understanding of the data flow between DT components. We argue that this approach enables practitioners to gain deeper insights into service-driven DTs, to better support their construction, evolution, and reporting.

%\noindent\paragraph*{Approach Overview}

In this article, we \textit{present the DTInsight tool and methodology, for continuously reporting a DT's conceptual architecture, its capabilities, and a selection of its behavior}. Our tool's source code is available at \cite{dtinsight}, with \cite{dtdfonto} containing the reporting framework ontology and our example modeled DT.

\begin{figure*}[htbp]
    \centering
    \includegraphics[width=\textwidth]{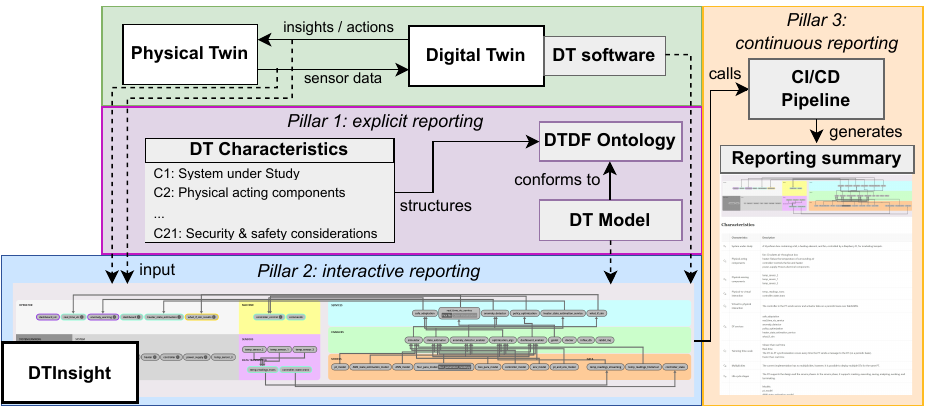}
    \caption{DTInsight monitors the communication between Physical and Digital Twin, reads the modeled DT, and generates a report.}
    \label{fig:architecture-diagram}
\end{figure*}

Our DTInsight approach has three pillars, as in Figure~\ref{fig:architecture-diagram}:

Pillar 1) \textit{Explicit reporting}, of the DT in the DTDF (Section~\ref{sec:dtdf}):
Explicit DT reporting is achieved by modeling the DT using the 21 characteristics from the DT Description Framework (DTDF)~\cite{gilSystematicReportingFramework2024}, in the Ontology Modeling Language (OML). We use openCAESAR \cite{elaasarOpenCAESARBalancingAgility2023}, an ontology development framework from NASA Jet Propulsion Laboratory (JPL). These DTDF characteristics are formalized in an OML \textit{vocabulary}, providing concepts for DT \textit{description models}.
%Section \ref{dtdf} presents the Digital Twin Description Framework as an ontology and illustrates its use through an incubator example.

Pillar 2) \textit{Interactive reporting}, through a game engine and system monitoring (Section~\ref{sec:architecture_methodology}): The interactive reporting uses description models conforming with the DTDF vocabulary to generate an interactive conceptual architecture of our DT, in a visualization called a \textit{DT constellation} \cite{Oakes2021improving}. Using the Godot Engine~\cite{godotengine}, we can represent the flow of data from the system's sensors with the appropriate communication libraries.

Pillar 3) \textit{Continuous reporting}, through integration into CI/CD (Section~\ref{sec:report_generation}): Finally, this visualization is integrated in a GitHub Actions CI/CD pipeline to create an interactive and up-to-date reporting page of the DT. This reporting page uses the description model of the DT to generate a conceptual architecture diagram, a summary table of the reporting characteristics, and an embedding of the interactive visualization.

%\paragraph*{Contributions and Structure}

%Section \ref{discussion} discusses the benefits and limitations of our approach, Section \ref{related-work} presents related work and Section \ref{conclusion} concludes. 

%\input{text/background}

\section{Explicit DT Description Framework Ontology} \label{sec:dtdf}

Previous work has found that DTs experience reports often lack crucial information for comparing and understanding the capabilities of said DT~\cite{Oakes2021improving,Oakes2023DigitalTwinDescription}. Therefore, we advocate for DTs' \textit{explicit reporting} through a systematic and consistent reporting framework composed of 21 characteristics to describe DTs~\cite{gilSystematicReportingFramework2024}. These reporting characteristics range from the twinning time-scale (C7), to fidelity and validity considerations (C14), technical implementation (C15), and security and safety considerations (C21).

We formalize this \textit{DT Description Framework} (DTDF) as an ontology, available at~\cite{dtdfonto}. An ontology is ``an explicit specification of a conceptualization''~\cite{gruber1993translation}. It defines classes, relations, functions, and axioms that make shared meaning machine-readable. For increased agility and rigor, we model the DTDF using OML in the OML Rosetta editor \cite{elaasarOpenCAESARBalancingAgility2023}. OML acts as a Domain Specific Language (DSL) layer above the well-known Web Ontology Language (OWL) to simplify ontology creation. It represents ontologies as \textit{vocabularies} (containing concepts, similar to a meta-model) and \textit{descriptions} (containing instances of those concepts). The DTDF vocabulary thus formalizes the 21 DT reporting characteristics and their ontological relationships.

Then, in an OML \textit{description model}, we model the incubator DT example \cite{feng2021incubator,gilSystematicReportingFramework2024}. The incubator's purpose is to maintain the temperature in its enclosure through the control of a heater and the monitoring of temperature sensors. The incubator DT can then visualize the incubator's behavior, optimize the control policy, etc. The incubator DT is proposed as a case for DT engineering as it has complex behavior in multiple domains (electrical, thermal, mechanical) yet is simple enough for pedagogical purposes~\cite{gomes2025digital}.

The DTDF vocabulary (excerpted in Listing~\ref{lst:dtdf_vocab_oml}) adopts a three-layered, service-oriented conceptual architecture for describing DTs. \textit{Models} and \textit{Data} are inputs to computational components called \textit{Enablers}, which process the models and data to enable \textit{Services} providing the DT's actions or insights.

\begin{figure}[tbh]
\begin{lstlisting}[language=oml, caption=DTDF ontology vocabulary model (DTDFVocab) excerpt in OML., label=lst:dtdf_vocab_oml]
// C6: Services
concept Service < DTComponent, TimeScaleThing
relation entity Provides [from Service to ProvidedThing forward provides reverse providedBy]
// C11: Enablers
concept Enabler < DTComponent
relation entity Enables [from Enabler to Service forward enables reverse enabledBy]
scalar property IsCommEnabler [domain Enabler range xsd:boolean functional]
// C10: Models/Data
aspect Input
concept Model < DTComponent, Input
concept Data < DTComponent, Input
relation entity InputTo [from Input to Enabler forward inputTo reverse hasInput]
relation entity DataInput [from DataTransmitted to Data forward asData reverse fromData]
// C20: Standardization
concept Standardization < base:DescribedThing
\end{lstlisting}
\end{figure}

The DTDF description imports the DTDFVocab to \textit{instantiate} its defined concepts. This allows each DT use case to reuse the vocabulary by creating instances tailored to its specific conceptual architecture. Listing~\ref{lst:dtdf_desc_oml} illustrates this approach through an example describing the incubator using the DTDF.

\begin{figure}[tbh]
\begin{lstlisting}[language=oml, caption=DTDF ontology incubator description model excerpt in OML., label=lst:dtdf_desc_oml]
// SERVICE EXAMPLE (C6)
instance what_if_sim : DTDFVocab:Service [DTDFVocab:provides what_if_sim_results DTDFVocab:atStage baseDesc:operation]
// ENABLER EXAMPLE (C11)
instance simulator : DTDFVocab:Enabler [DTDFVocab:enables what_if_sim]
// MODEL EXAMPLE (C10)
instance controller_model : DTDFVocab:Model [DTDFVocab:inputTo simulator DTDFVocab:inputTo state_estimator DTDFVocab:inputTo optimization_algs]
// DESCRIBED CHARACTERISTIC EXAMPLE (C20)
instance standardization : DTDFVocab:Standardization [base:desc "Communication is carried out using AMQP standard via RabbitMQ. Behavioral models have been produced following the FMI standard version 2."]
\end{lstlisting}
\end{figure}

\section{Interactive DTInsight Tool} \label{sec:architecture_methodology}

This section describes how DTInsight loads the user's DT as described in the DTDF, and generates an interactive visualization. The objective is to unify the structural and behavioral descriptions of the DT and extend the user's reporting view to incorporate both reporting \textit{and} behavioral insights. 

\paragraph{Technical Details}

DTInsight is built in the Godot Engine \cite{godotengine}, an open-source game engine with active development and community. It is well-suited to our needs as it a) is MIT-licensed, b) supports both 2D and 3D, with advanced UI creation capabilities, c) exports to web and all desktop platforms, and d) focuses on ease-of-use. By using the .NET version of Godot, we can also subscribe to RabbitMQ \cite{rabbitmq} message broker queues through available C\# libraries, enabling us to capture data flowing from the real or simulated incubator system and display it in the visualization.

The DTDF ontology is served by the Apache Jena Fuseki server in Rosetta~\cite{elaasarOpenCAESARBalancingAgility2023}. We use \textit{queries} to retrieve the details of the DT instance described in DTDF via SPARQL, which is a standard query language for retrieving and manipulating data stored in Resource Description Framework (RDF) format. Users interact with the Fuseki server by sending HTTP GET requests to its endpoint, allowing them to retrieve and explore DTDF ontological data represented in RDF. This functionality facilitates the querying of complex datasets and the extraction of specific relationships and entities, enabling efficient data retrieval within semantic web applications.

% Our running example of a simple incubator is used portray this ontology in Listing~\ref{lst:dtdf_oml}

From these query results, DTInsight creates a DT visualization  termed a \textit{constellation}, following the approach suggested in \cite{Oakes2021improving}. A \textit{DT constellation} is a conceptual architecture that represents the flow of data between Physical Twin (PT), DT, and within the DT. The DT is thus viewed ``as an agglomeration of all related models, data, enablers, and [services] that are used in the DT activities''~\cite{Oakes2021improving}.

Note that while other ontology visualizations exist~\cite{dudavs2018ontology}, these often use conventional class- or graph-based visualization methods which are not suited for DTs and their rich semantic structure.  Instead, we see that this constellation view can improve stakeholder collaboration in DT engineering~\cite{goffi2025brewing}.

We make the DT constellation \textit{interactive}, to unlock the filtering and presenting of relevant and detailed information to the user \cite{stepanekInteractiveDiagramsSoftware2024a}, and \textit{behavioral}, by reflecting the incoming data flow of sensor information. These properties motivate the choice of Godot to build a dynamic visualization. 
%, and are detailed in Section \ref{fig:user-interaction}.

%The RabbitMQ connection parameters are themselves retrieved from the ontology server.

\paragraph{User Interaction with DTInsight} \label{fig:user-interaction}

\begin{figure}[ht]
    \centering
    \includegraphics[width=\linewidth]{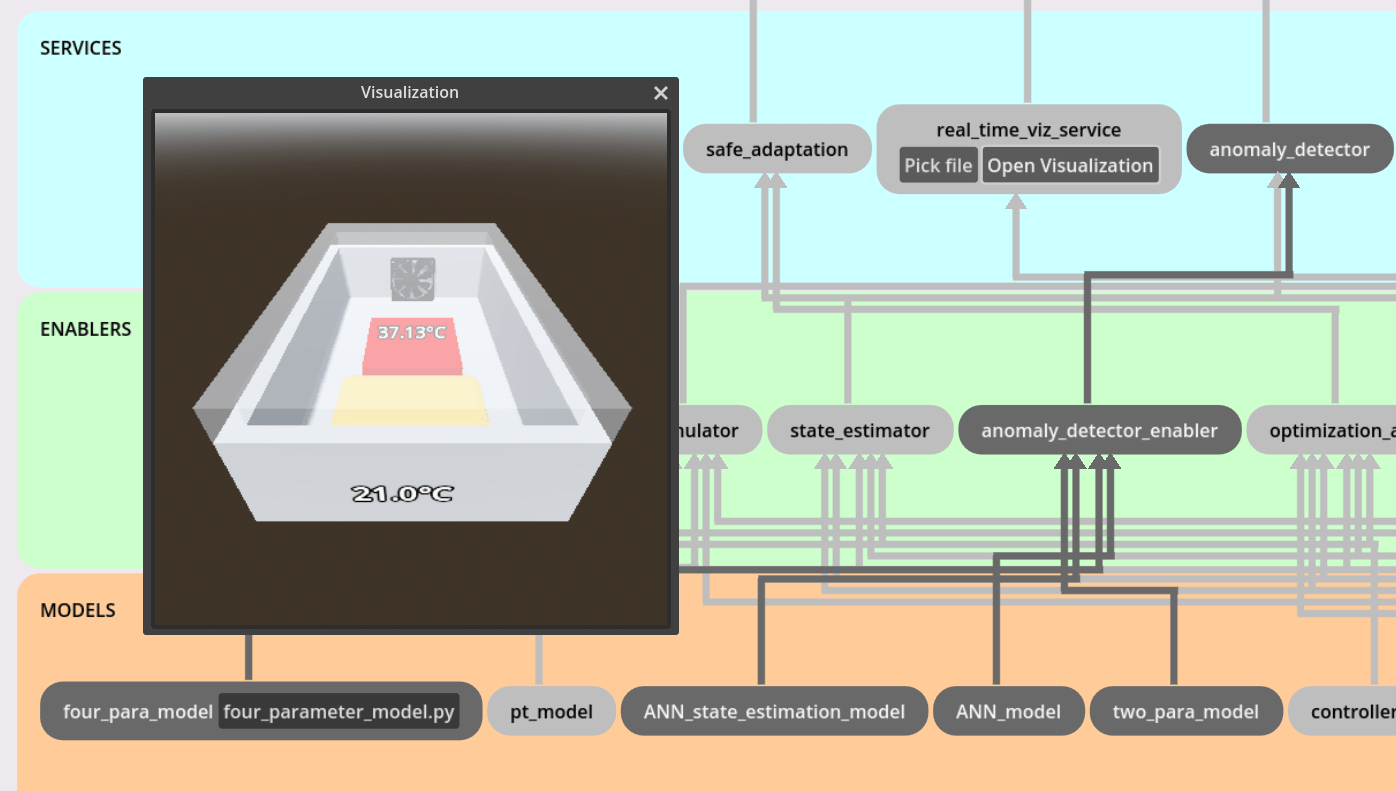}
    \caption{DT constellation interactivity: highlight DT component dependencies and display an up-to-date 3D visualization of the system}
    \label{fig:dtinsight-architecture}
\end{figure}

DTInsight offers users various features to gain insights into the system. Figure~\ref{fig:architecture-diagram} (DTInsight) illustrates the DT system’s composition, structured into two primary sections. The left side represents the five PT components: \textit{Operator}, \textit{Machine}, \textit{System Environment}, \textit{System}, and \textit{Sensors/Data Transmission}. The right side (from bottom-to-top) illustrates the three capability categories of the DT: \textit{Models/Data}, \textit{Enablers}, and \textit{Services}. One-directional arrows connect the components, indicating their inter-dependencies and data flow relationships.

Figure~\ref{fig:dtinsight-architecture} shows DT constellation \textit{interactivity}: hovering the mouse or clicking over DT components highlights connected components and their data flows. This provides the user with a \textit{structural} understanding of the DT by allowing them to easily explore its components as they are formally described in ontologies. It can help them find components that depend on each other, and navigate backward or forward in the layered and service-based DT conceptual architecture.

\textit{Additional Interaction:} If the user has connected the software folder containing the DT code, they can also click on a component to see its associated script. If they have connected RabbitMQ, they can click on a sensor component to see a graph of its real-time data.

And finally, the user can pick an external visualization file (a Godot resource pack) to open a 3D visualization of the DT in a pop-up that reflects incoming data from sensors. In the example incubator visualization presented in Figure~\ref{fig:dtinsight-architecture}, data labels are updated to reflect the temperature in real-time and the heater color turns red when heating. This strengthens the structural and behavioral descriptions, while keeping the 3D visualization decoupled from the conceptual architecture by being packed into an external file.

\section{Continuous Report Generation} \label{sec:report_generation}

With DTInsight, we are targeting both DT experts and non-technical personnel who benefit from \textit{continuous reporting} to stay updated on the evolution of the DT as it is being developed. From a CI/CD pipeline, currently implemented using GitHub Actions, we can automatically generate a characteristics table and visualizations\footnote{An example page for the incubator DT is found here: \url{https://oakeslabmtl.github.io/DTDF/}} from every commit to the repo containing the DT code and description model.

Thus, the reporting page deployment takes three steps:

\begin{enumerate}
    \item Modeling the DT in the DTDF (in OML/OWL/RDF)
    \item Setting up the deployment of the static website
    \item Running the CI/CD workflow to create the report page
\end{enumerate}

The first step is that the user updates the description model of their DT or system using an ontology editing tool such as openCAESAR Rosetta, which is then committed to a Git repository.
Upon committing, the reporting pipeline loads the ontology in a Fuseki server and runs the DTInsight tool in a Linux xvfb display server. We trigger its query in Godot via an HTTP request. It outputs the HTML characteristics table, the screenshot of the conceptual architecture, and the YAML file describing it. The web-export of DTInsight is embedded as an iframe into the reporting page, which is then deployed on the web as a static website using Hugo~\cite{gohugo}.

Thus, as the underlying ontology of the DT evolves, the reporting page reflecting the DT characteristics and conceptual architecture is automatically updated to mirror these changes.

\section{Related Work} \label{related-work}

\paragraph{DT Architecture} Dalibor \textit{et al.} employ DSLs to generate interactive DT cockpits that integrate both the internal DT infrastructure and a monitoring frontend~\cite{daliborModelDrivenArchitectureInteractive2020}. A similar low-code approach is taken by De Sanctis \textit{et al.} in the context of smart cities~\cite{desanctisLowcodeAssessmentPlatform2025}.
However, they do not adopt the ontological and service-oriented approach resulting in the DT constellation view \cite{oakesOntologicalServiceDrivenEngineering2024}, nor do they report infrastructure evolution within a continuous CI/CD pipeline.
Carrion \textit{et al.} recently proposed a conceptual Entity-Relationship Digital Twin (ERDT) model to represent the PT, DT, and their interconnections~\cite{carrionConceptual2025}. Information retrieval within this model is facilitated through DSL-defined queries \cite{CarrionQueryViews2025,carrionQueryingDigitalTwin2025a}. In contrast, our approach leverages an ontological framework that enables formal semantics, logical inference, and automated reasoning, using SPARQL for queries.
Our work resembles Software Architecture Reconstruction (SAR) explored by Ducasse  \textit{et al.}~\cite{ducasseSoftwareArchitectureReconstruction2009}, but applied to DTs. Unlike traditional \textit{reverse architecting} from source code, we perform \textit{manual SAR} by leveraging user-defined ontologies to construct a \textit{conceptual and architectural view}, utilizing the DT constellation’s \textit{reporting style}.
Ozkaya \textit{et al.} proposed a modeling language for DT architecture using C4 (Context, Containers, Components, Code) \cite{ozkaya2024towards}. In contrast, our work focuses on reporting DT architecture components in the DTDF, using ontologies as ground truth.

\paragraph{DT Reporting} The lack of adequate documentation in DT systems and the frequent divergence between actual and original design is highlighted by Gunasekaran \textit{et al.}~\cite{gunasekaran2025behavioralanalysis}. To address this, their work leverages logging to analyze the behavior of evolving DT systems. In contrast, our approach focuses on simplifying the generation of continuous software documentation through DTDF reporting driven by expert knowledge.
Yahouni \textit{et al.} propose a reporting system for decision-making actors during manufacturing. It uses a Multi-Agent System (MAS) to gather user needs, extract data, calculate Key Performance Indicators (KPIs), and generate reports~\cite{yahouniSmartReportingFramework2021}.  In contrast, our approach is centered on generating \textit{conceptual} DT reports, focusing on their architecture and characteristics. Uhlenkamp \textit{et al.} propose a DT maturity assessment tool\footnote{\url{https://dt-maturity.eu/}}, allowing the user to assign values to their assessment categories through a web-based form and obtain a maturity score~\cite{uhlenkamp2022digital}. By comparison, we focus on the conceptual DT architecture, describing DTs using free-form text for most characteristics.

\paragraph{Software Visualization} Langelier \textit{et al.} used visualization to analyze large scale-systems \cite{langelier2005visualization}. Similarly, Cerny \textit{et al.} explored microservice architecture visualization techniques for static and dynamic SAR \cite{cerny2022microservice,cernyStaticCodeAnalysis2024a}. Meanwhile, Antoniazzi \textit{et al.} focused on RDF graph visualization \cite{antoniazziGraphVisualization2018}. We bring those works into SAR for (micro)service-based DTs using the DTDF ontology, structuring its RDF graph as a DT constellation visualization and making it behavioral by integrating data flow.

% SIT is publicly available at GitHub - osslab-pku/SIT: SIT: An accurate, compliant SBOM generator with incremental construction, and a demonstration video can be found at SIT: An accurate, compliant SBOM generator with incremental construction. \url{https://github.com/osslab-pku/SIT}

\section{Conclusion} \label{conclusion}

DTInsight's primary benefit is enhancing communication with stakeholders, which is achieved through a threefold contribution to DT reporting: making it \textit{explicit} through the DTDF ontology, \textit{interactive} via the structural and behavioral conceptual architecture, and \textit{continuous} through reporting page generation. This provides a domain-specific view on DTs, building on prior work on a DT reporting framework \cite{Oakes2021improving,gilSystematicReportingFramework2024}. Furthermore, the integration of scripts directly into the DT constellation and the inclusion of real-time data in graphs and 3D visualizations promote a practice of interactive monitoring. % This was demonstrated to be appreciated by multiple DT practitioners \todo{ask Claudio and others}.

\paragraph*{Limitations}
However, there are some limitations to consider. The system presents a high-level conceptual visualization of the DT architecture, but it requires modeling by a DT expert that has an overarching understanding of the DT. Moreover, the visualization itself could be made more readable and offer a denser representation of information. Going deeper into examining each component of the DT would also promote  observability~\cite{majors2022observability}. On the technical side, the system currently supports only one message broker (RabbitMQ), which limits its flexibility. Additionally, the reliance on manually written ontologies is a limitation, though this could be improved by using techniques like Large Language Models (LLMs) for automatic ontology generation, or by incorporating visual interfaces to offer a creation GUI. Another limitation is that the current tool only supports the representation of a single DT at a time, but this could be expanded to represent systems-of-systems \cite{adesanya2025systems}. Furthermore, the system does not support real-time changes to represent self-reconfiguring DTs~\cite{kamburjan2022digital}. This aligns with research on human/DT interaction with intelligent and adaptive interfaces  capable of evolving in real time and adapting to both the operator’s context and needs \cite{palmerNeedSymbioticInterface2023}.

\paragraph*{Future work}
We are currently pursuing editing the description model directly from a visual interface, assisted by integrated LLMs. We are also looking at exposing more information about each component in the DT, such as its current lifecycle stage, operational state, and visualizing simulation results over time.

\newpage

% \section*{Acknowledgment} 
% This work supported by NSERC grant RGPIN-2024-05622.

\bibliographystyle{IEEEtran}
\bibliography{bibliography}

\end{document}